\documentclass{aastex62}
\shorttitle{Evidence of Black Hole Burning in $\omega$ Centauri}
\shortauthors{Cheng et al.}
\begin{document}
\title{Exploring the Mass Segregation Effect of X-ray Sources in Globular Clusters. IV. Evidence of Black Hole Burning in $\omega$ Centauri}
\correspondingauthor{Zhongqun Cheng}
\email{chengzq@whu.edu.cn}
\author[0000-0002-9983-8609]{Zhongqun Cheng}
\affiliation{School of Physics and Technology, Wuhan University, Wuhan 430072, People's Republic of China}
\affiliation{WHU-NAOC Joint Center for Astronomy, Wuhan University, Wuhan 430072, People's Republic of China}

\author[0000-0003-0355-6437]{Zhiyuan Li}
\affiliation{School of Astronomy and Space Science, Nanjing University, Nanjing 210023, People's Republic of China} 
\affiliation{Key Laboratory of Modern Astronomy and Astrophysics (Nanjing University), Ministry of Education, Nanjing 210023, People's Republic of China}

\author[0000-0003-3901-8403]{Wei Wang}
\affiliation{School of Physics and Technology, Wuhan University, Wuhan 430072, People's Republic of China}
\affiliation{WHU-NAOC Joint Center for Astronomy, Wuhan University, Wuhan 430072, People's Republic of China}

\author[0000-0002-0584-8145]{Xiangdong Li}
\affiliation{School of Astronomy and Space Science, Nanjing University, Nanjing 210023, People's Republic of China} 
\affiliation{Key Laboratory of Modern Astronomy and Astrophysics (Nanjing University), Ministry of Education, Nanjing 210023, People's Republic of China}

\author[0000-0002-3614-1070]{Xiaojie Xu}
\affiliation{School of Astronomy and Space Science, Nanjing University, Nanjing 210023, People's Republic of China} 
\affiliation{Key Laboratory of Modern Astronomy and Astrophysics (Nanjing University), Ministry of Education, Nanjing 210023, People's Republic of China}

\begin{abstract}

Using X-ray sources as sensitive probes of stellar dynamical interactions in globular clusters (GCs), we study the mass segregation and binary burning processes in $\omega$ Cen. We show that the mass segregation of X-ray sources is quenched in $\omega$ Cen, while the X-ray source abundance of $\omega$ Cen is much smaller than other GCs, and the binary hardness ratio (defined as $L_{\rm X}/(L_{\rm K}f_{b})$, with $f_{b}$ the binary fraction, $L_{\rm X}$ and $L_{\rm K}$ the cumulative X-ray and K band luminosity of GCs, respectively) of $\omega$ Cen is located far below the $L_{\rm X}/(L_{\rm K}f_{b})-\sigma_{c}$ correlation line of the dynamically normal GCs. These evidences suggest that the binary burning processes are highly suppressed in $\omega$ Cen, and other ``heating mechanisms", very likely a black hole subsystem (BHS), are essential in the dynamical evolution of $\omega$ Cen. Through the ``black hole burning" processes (i.e., dynamical hardening of the BH binaries), the BHS is dominating the energy production of $\omega$ Cen, which also makes $\omega$ Cen a promising factory of gravitational-wave sources in the Galaxy.

\end{abstract}


\section{Introduction}
It is well-known that the dynamical evolution of self-gravitating systems is unstable if without an internal energy source. Two representative systems, stars and GCs, are natural laboratories to verify this theory. 
Indeed, both stars and GCs are highly evolved (i.e., with the dynamical evolution Kelvin-Helmholtz timescale or two-body relaxation timescale much smaller than their age) and a ``heating mechanism" is essential in maintaining their thermal (or quasi-thermal) equilibrium states and avoiding core collapse \citep{heggie2003}. 
Moreover, stars and GCs are also very similar to each other in the aspect of evolution: both systems are subtly balanced by the production of energy in the core and the outflow of energy from the system, which suggest that {\it it is the structure of the whole system rather than only the central core that controls the energy production rate of a star (or GC)}. 
For stars, this principle was first summarized by \citet{eddington1926}, and is regarded as one of the fundamental principles of stellar evolution. While the same principle in GCs was first realized by \citet{henon1975}, although the nature of the internal energy source was not very clear at that time.

We now know that binary is the pivotal ingredient of internal energy source in GCs. Through dynamical encounters, hard binaries (with bound energy $|E_{b}|$ larger than the average stellar kinetic energy $E_{k}$ of stars in GCs) tend to transfer energy to fly-by stars and ``heat" the cluster (i.e., the binary burning processes), which is sufficient to delay, halt, or even reverse the core-collapse of GCs \citep{heggie2003,fregeau2003}.   
Meanwhile, the surviving main sequence (MS) binaries will become more and more compact after the dynamical encounters \citep{hills1975,heggie1975}, which may lead to the formation of many exotic objects in GCs, such as low-mass X-ray binaries (LMXBs; e.g., through exchange encounters; \citealp{rasio2000,ivanova2010,giesler2018,kremer2018a}), millisecond pulsars (MSPs; the offspring of LMXBs; \citealp{ivanova2008,ye2019}), cataclysmic variables (CVs; \citealp{ivanova2006,shara2006,belloni2016,hong2017,belloni2017,belloni2019}), coronal active binaries (ABs) and blue straggler stars (BSS; \citealp{fregeau2004,chatterjee2013}). 
According to the H{\'e}non principle, there should be a global parameter, which may connect the cluster structure and binary burning products in GCs. 

In fact, this parameter is the dynamical encounter factor (i.e., $\Gamma$, defined as an integral over the cluster volume $\Gamma \propto \int \rho^{2}/\sigma$, with $\rho$ the stellar density and $\sigma$ the velocity dispersion), which was proposed by \citet{verbunt1987} to solve the formation problem\footnote{The abundances of LMXBs and MSPs are more than $\sim 100$ times higher in GCs than in the Galactic field \citep{clark1975,katz1975,camilo2005,ransom2008}.} of LMXBs and MSPs in GCs. Applying $\Gamma$ to weak X-ray sources (mainly CVs and ABs), \citet{pooley2003} suggested that most weak X-ray sources are dynamically originated in GCs, by finding a strong correlation between the X-ray source counts ($N$) and $\Gamma$ in a dozen GCs. However, using the GC cumulative X-ray luminosity ($L_{X}$) as a proxy of $N$, \citet{cheng2018a} demonstrated that $L_{X}$, $\Gamma$ and cluster mass are highly correlated with each other in GCs, and most GCs have a slightly lower X-ray emissivity (i.e., $L_{X}$ per unit stellar mass) thus a slightly lower abundance of CVs and ABs than the Galactic field, a trend opposite to LMXBs and MSPs. This evidence indicates that the formation of CVs and ABs is suppressed, at least, in low density GCs \citep{heinke2020}, and many primordial binaries have been dynamically disrupted before they could otherwise evolve into weak X-ray sources in GCs \citep{cheng2018a}.

The lower abundance of CVs and ABs in most GCs than in the Galactic field is a challenge to our understanding of  binary burning processes as an internal energy source in GCs. Moreover, if most of the MS binaries were dynamically disrupted rather than transformed into weak X-ray sources, the Hills-Heggie law suggests that they will serve as a ``cooling mechanism" in GCs \citep{hills1975,heggie1975}, a process further increasing the energy budget of cluster dynamical evolution. Taking these aspects together, it is reasonable to infer that, besides the binary burning processes, there are other alternative ``heating mechanisms" in GCs. 

Indeed, simulations have shown that BHs may play a fundamental role in the dynamical evolution of GCs. Due to the mass segregation effect, BHs are expected to concentrate in the cluster center quickly \citep{spitzer1969}, where they may form a high-density subsystem and ``heat" the host cluster through BH burning\footnote{Following the suggestion of \citet{kremer2020a}, we call the dynamical hardening of BH binaries as BH burning in this paper, which is distinct from the binary burning because the products are invisible in electromagnetic waves.} processes \citep{merritt2004, mackey2007, mackey2008, breen2013, arcasedda2018}.
Contrary to the traditional idea that the BHS is dynamically decoupled from the host cluster and would lead to the ejection of most BHs within a few gigayears \citep{sigurdsson1993,kulkarni1993}, modern state-of-the-art simulations of GCs suggest that there is thermal contact between BHs and stars \citep{morscher2013,morscher2015,wang2016,kremer2019}, which dramatically increases the timescale of BH evaporation in GCs. As a result, the BHS may serve as a long-lasting energy source in GC evolution, and many exotic objects, including gravitational wave sources \citep{portegies2000, downing2010, banerjee2010, samsing2014,rodriguez2015,rodriguez2016,askar2017,hong2018,antonini2020} and intermediate-mass black holes\footnote{Once the IMBH is formed, it also may serve as a long-lasting internal energy source in GCs \citep{baumgardt2004, trenti2007, gill2008, trenti2013}.} (IMBHs; \citealp{miller2002,giersz2015,antonini2019}), could be created through the dynamical hardening and merger of BH binaries in GCs.

According to the H{\'e}non principle, the co-evolution of BHS and host clusters may lead to remarkable influence on the structure of GCs, especially in the cluster core.
For example, compared with clusters with few BHs, simulations show that GCs with a large number of BHs are more likely to have a large core and low central density \citep{merritt2004, hurley2007, chatterjee2017, arcasedda2018, askar2018,kremer2018b,kremer2019}, which may quench or slow down the mass segregation of normal stars in GCs \citep{alessandrini2016,peuten2016,weatherford2018}. On the other hand, with a BHS occupies the cluster core and dominates the production of energy, the binary burning processes will be highly suppressed in GCs, and GCs host a large number of BHs are more likely to have a lower formation efficiency of LMXBs, MSPs, CVs, ABs and BSS \citep{ye2019,kremer2020b}. As a result, these phenomena may offer us useful hints to evaluate the population of BHs in GCs.

Observationally, there are a growing number of BH candidates that have been identified in Galactic and extragalactic GCs \citep{maccarone2007, strader2012, chomiuk2013, millerjonse2015, bahramian2017, shishkovsky2018, giesers2018, giesers2019, zhao2020}. Some clusters, such as M 22 \citep{strader2012} and NGC 3201 \citep{giesers2018, giesers2019}, are found to host more than one BH candidate, strongly supporting the existence of a BHS in these systems. Based on the influence of BHS on the cluster structure, many authors have studied the existence of BHS and estimated the populations of BHs in many Galactic GCs \citep{arcasedda2018, askar2018, askar2019, weatherford2020, askar2020}. However, the suppression effect of GC binary burning processes by BHS is rarely explored in the literature, and it remains an open question how a BHS would influence the energy production of the host cluster.

In the previous work, we have demonstrated that the X-ray sources are sensitive probes of stellar dynamical interactions in GCs. For example, by comparing the abundance ratio of X-ray-emitting close binaries to MS binaries in star clusters, we have studied binary burning processes and tested Hills-Heggie law within 30 GCs \citep{cheng2018b}. On the other hand, we found significant distribution dips for X-ray sources in 47 Tuc, which is consistent with the mass segregation delay of heavy objects in GCs \citep{cheng2019a}. Using the mass segregation effect of X-ray sources as an indicator of cluster dynamical age, we also show that there is evident acceleration effect for cluster dynamical evolution by tidal stripping in Terzan 5 \citep{cheng2019b} and M28 \citep{cheng2020}. Besides, we found an abnormal deficiency of X-ray sources in the central region ($R\lesssim1.5\arcmin$) of M28 with respect to its outskirts, which indicates that M28 has suffered an early phase of primordial binary disruption within its central region, and that BHS may play an important role in creating and maintaining this peculiar distribution of X-ray sources in M28 \citep{cheng2020}. 

In this work, we explore the existence of BH burning in $\omega$ Cen, by using the weak X-ray sources as sensitive probes of mass segregation effect and binary burning processes in this cluster.
With the largest mass among Galactic GCs ($M=3.55\pm 0.03\times 10^{6}\, M_{\odot}$; \citealp{baumgardt2018}), and a high central velocity dispersion ($\sigma_{c}=19.09^{+1.46}_{-1.29}\, \rm km\, s^{-1}$; \citealp{baumgardt2019a}), $\omega$ Cen is one of the prime targets to search for candidate IMBHs and BHS. In fact, many authors have argued that $\omega$ Cen may host an IMBH, based on the fitting of cluster surface brightness and velocity dispersion profile \citep{noyola2008, jalali2012, baumgardt2017}. However, the IMBH scenario is challenged by the presence of a BHS in $\omega$ Cen, which is thought to have a similar effect in shaping the structure of this cluster \citep{vandermarel2010,zocchi2019}.
Recently, using the updated {\it HST}/ACS proper motion catalogue of \citet{bellini2017a}, \citet{baumgardt2019a} have ruled out the existence of a massive IMBH in $\omega$ Cen. Instead, they argue that all the data of $\omega$ Cen can be fitted by a BHS model that contains $4.6\%$ of the mass of $\omega$ Cen. 

The remainder of this paper is organized as follows. In Section 2, we describe the X-ray data analysis and creation of the X-ray source catalog. We study the cluster mass segregation effect and binary burning processes in Section 3, and explore its relation to the BHS in Section 4. Our main conclusions are outlined in Section 5.

\section{X-ray Data Analysis}
\subsection{{\it Chandra} Observations and Data Preparation}

So far, there are 4 archival {\it Chandra} AICS-I observations for $\omega$ Cen. The 2001 (ObsID 653 and 1519) and 2012 (ObsID 13726 and 13727) data sets have been analysed by \citet{haggard2009} and \citet{henleywillis2018}, respectively. In order to detect the faintest sources in $\omega$ Cen, all observations were used in our X-ray data analysis, which amount to a total effective exposure of $\sim 290$ ks in the central region of $\omega$ Cen. 
Following the standard procedure\footnote{http://cxc.harvard.edu/ciao}, we used the {\it Chandra} Interactive Analysis Observations (CIAO, version 4.11) and the {\it Chandra} Calibration Database (version 4.8.4) to reprocess the data.
By setting the ObsID 13726 as the reference frame, we corrected the relative astrometry among the 4 observations, and then merged them into master event files using the CIAO tool {\it merge\_obs}. Three groups of images have been created in soft (0.5--2 keV), hard (2--8 keV), and full (0.5--8 keV) bands and with a bin size of 0.5, 1, and 2 pixels, respectively.
These images are used for detecting X-ray sources in this work.

\subsection{Sources Detection and Sensitivity}

To obtain a reliable source catalog, we employed a two-stage approach to build the X-ray sources list in $\omega$ Cen. First, we ran {\it wavdetect} within the field of view (FoV) of the 9 combined images, using the ``$\sqrt{2}$ sequence" wavelet scales and aggressive false-positive probability thresholds to find weak X-ray sources. If possible X-ray sources were missed by the {\it wavdetect} script, we also inspected the merged images visually and added them into the detection source list. The {\it wavdetect} results were combined into a master source list, which resulted in a candidate list of 344 sources. 
Then, we utilized the ACIS Extract (AE; \citealp{broos2010}) package to extract and filter the candidate source list. The AE binomial no-source probability ($P_{\rm B}$) parameter was adopted to evaluate the significance of X-ray sources.
As demonstrated in the case of 47 Tucanae, we set a more stringent threshold value of $P_{\rm B}\lesssim 1\times 10^{-3}$ for validated X-ray sources, which was proved to be helpful in optimizing the source completeness and reliability \citep{cheng2019a}.
On the other hand, the threshold value of $P_{\rm B}$ also offers us a criterion to calculate the limiting detection sensitivity across the FoV of $\omega$ Cen. The resulting sensitivity maps were found to be very useful in estimating the contribution of cosmic X-ray background (CXB) sources and Galactic interlopers \citep{cheng2019b, cheng2020}.

\begin{figure}[htb!]
\centering
\includegraphics[angle=0,origin=br,height=0.4\textheight, width=0.51\textwidth]{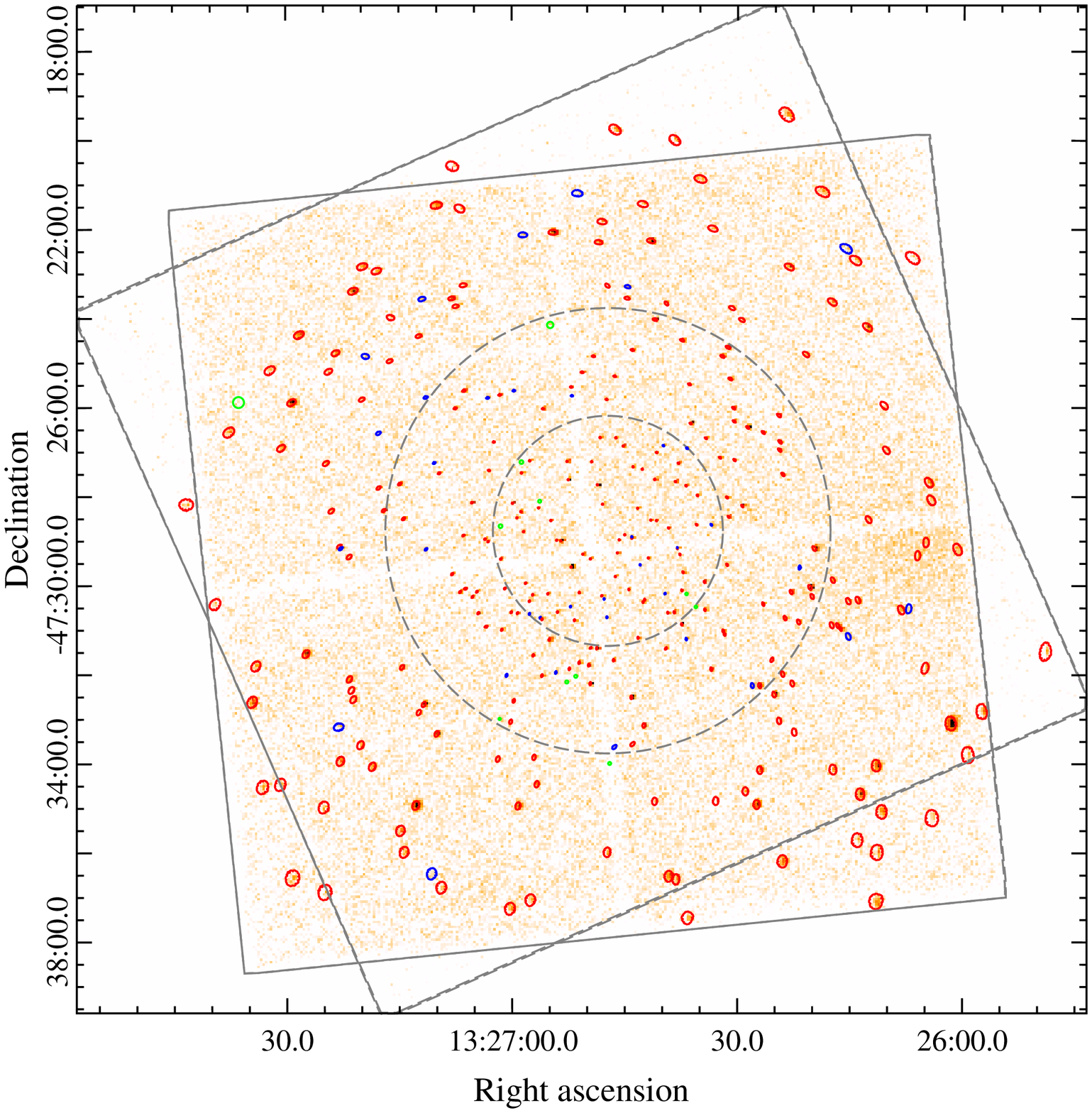}\includegraphics[angle=0,origin=br,height=0.4\textheight, width=0.53\textwidth]{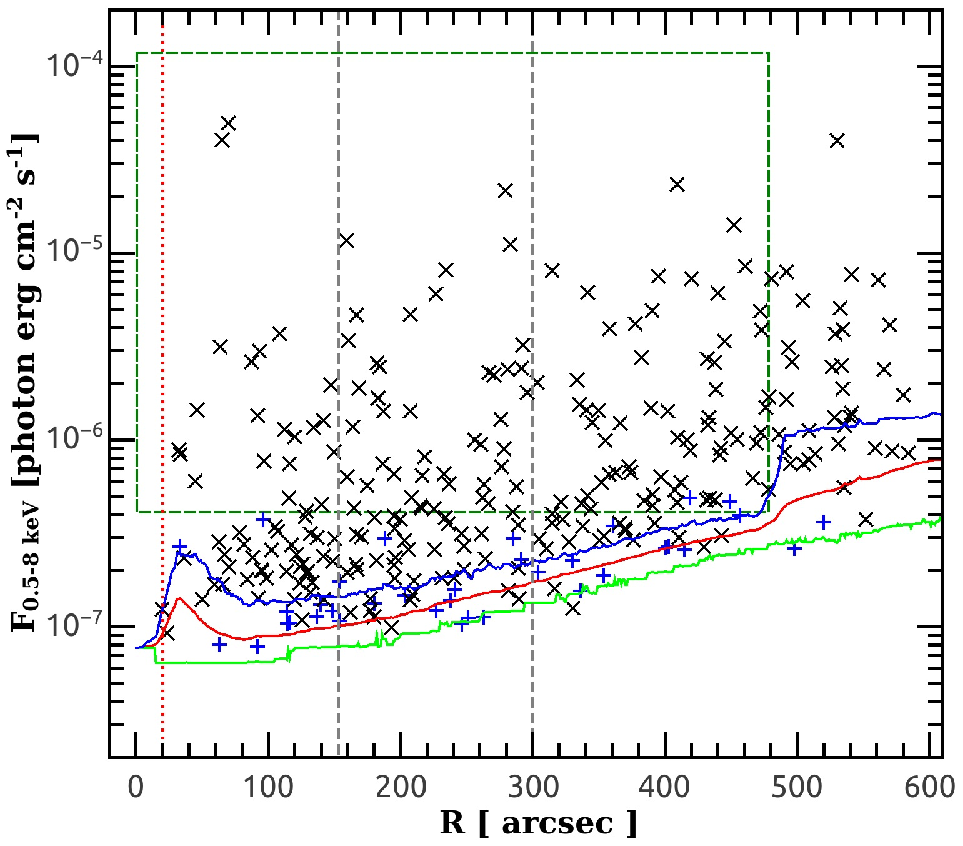}
\linespread{0.7}
\caption{Left: The 0.5-8 keV {\it Chandra} merged image of $\omega$ Cen. The image is smoothed with a Gaussian kernel with a radius of 3 pixels. FoVs of the 2001 (ObsID 653 and 1519) and 2012 (ObsID 13726 and 13727) data sets are indicated by dashed boxes. Sources detected by \citet{haggard2009} and \citet{henleywillis2018} are marked with red ellipses and green circles, while new detections of this work are shown as blue ellipses. The gray dashed circles represent the cluster core radius ($R_{c}=153\arcsec$; \citealp{ferraro2006}) and half-light radius ($R_{h}=5\arcmin$; \citealp{harris1996}). Right: 0.5-8 keV photon flux as a function of projected distance from the cluster center. New detections of this work are displayed as blue pluses. The color-coded solid curves represent the median (red), minimum (green), and maximum (blue) limiting detection count rates at corresponding radial bins, respectively. The sensitivity peaks at $R\sim 40\arcsec$ are caused by the CCD gaps of the {\it Chandra} ACIS-I array. The cluster core and half-light radius are shown as gray dashed vertical lines, while the radius of influence of the possible massive IMBH ($R_{roi}=20\arcsec$; \citealp{baumgardt2019a}) is marked as a red dotted vertical line. X-ray sources selected by the olive dashed box are shown in Figure-\ref{fig:3} (a), which have the advantage of unbiased detection.   \label{fig:1}}
\end{figure}

Following the procedures presented in \citet{cheng2019a}, we checked the source validation and calculated the detection sensitivity maps within the {\it Chandra} FoV of $\omega$ Cen. Our final source catalog contains 300 X-ray sources, with $P_{\rm B}\lesssim 1\times 10^{-3}$ in either the full, soft or hard energy band. The locations of the X-ray sources are indicated by ellipses in the left panel of Figure-\ref{fig:1}. As a comparison, we also highlight the 275 X-ray sources identified by \citet{haggard2009} and \citet{henleywillis2018} as red and green ellipses in the figure. We reidentified those sources in red (264) but failed to identify those in green (11) based on our source pruning criterion. With a larger effective exposure than \citet{haggard2009} and \citet{henleywillis2018}, 36 sources (blue ellipses) are newly detected and tabulated in our catalog. We highlight these new detections as blue pluses in the right panel of Figure-\ref{fig:1}. It is evident that almost all of them are located above the minimum sensitivity line.

\subsection{Source Properties}

To derive the X-ray source properties, we utilized the AE package to extract the source photometry. The source and background spectra were first extracted from each of the 4 observations separately, then we merged the extraction results into composite event lists, spectra, light curves, response matrices and effective area files for each source through the ``MERGE\_OBSERVATIONS" procedure of AE.  
Aperture-corrected net source counts and photon fluxes are derived in soft, hard and full band, respectively. Among the 300 X-ray sources, 124 of them have net counts greater than $\sim 30$ in the full band. Using the AE automated spectral fitting script, we modeled their spectra with an absorbed power-law spectrum. In all cases the neutral hydrogen column density ($N_{\rm H}$) is constrained to not less than $N_{\rm H}=7.0\times10^{20}\,\rm cm^{-2}$, calculated from the color excess $E(B-V)$ of $\omega$ Cen \citep{harris1996}. 
For sources with net counts less than $\sim 30$, we converted their net count rates into unabsorbed energy fluxes, by using the AE-generated merged spectral response files and assuming a power-law model with fixed photon index ($\Gamma=1.4$) and absorption column density ($N_{\rm H}=7.0\times10^{20}\,\rm cm^{-2}$). 
Finally, we collated the source extraction and spectral fitting results into a main X-ray source catalog, with source labels sorted by their R.A.. Assuming a distance of 5.2 kpc for $\omega$ Cen \citep{harris1996}, we calculated for each X-ray source an unabsorbed luminosity in the soft, hard and full bands, respectively. For each source, we also computed a projected distance from the cluster optical center ($\alpha =13^{h}26^{m}47^{s}.24$ and $\delta =-47\arcdeg 28\arcmin 46.45\arcsec$; \citealp{anderson2010}). The final catalog of point sources is presented in Table 1 for the sake of future reference.

\section{Analysis: Source Raidal Distribution and Binary Burning Processes}

\begin{figure}[htb!]
\centering
\includegraphics[angle=0,origin=br,height=0.35\textheight, width=0.51\textwidth]{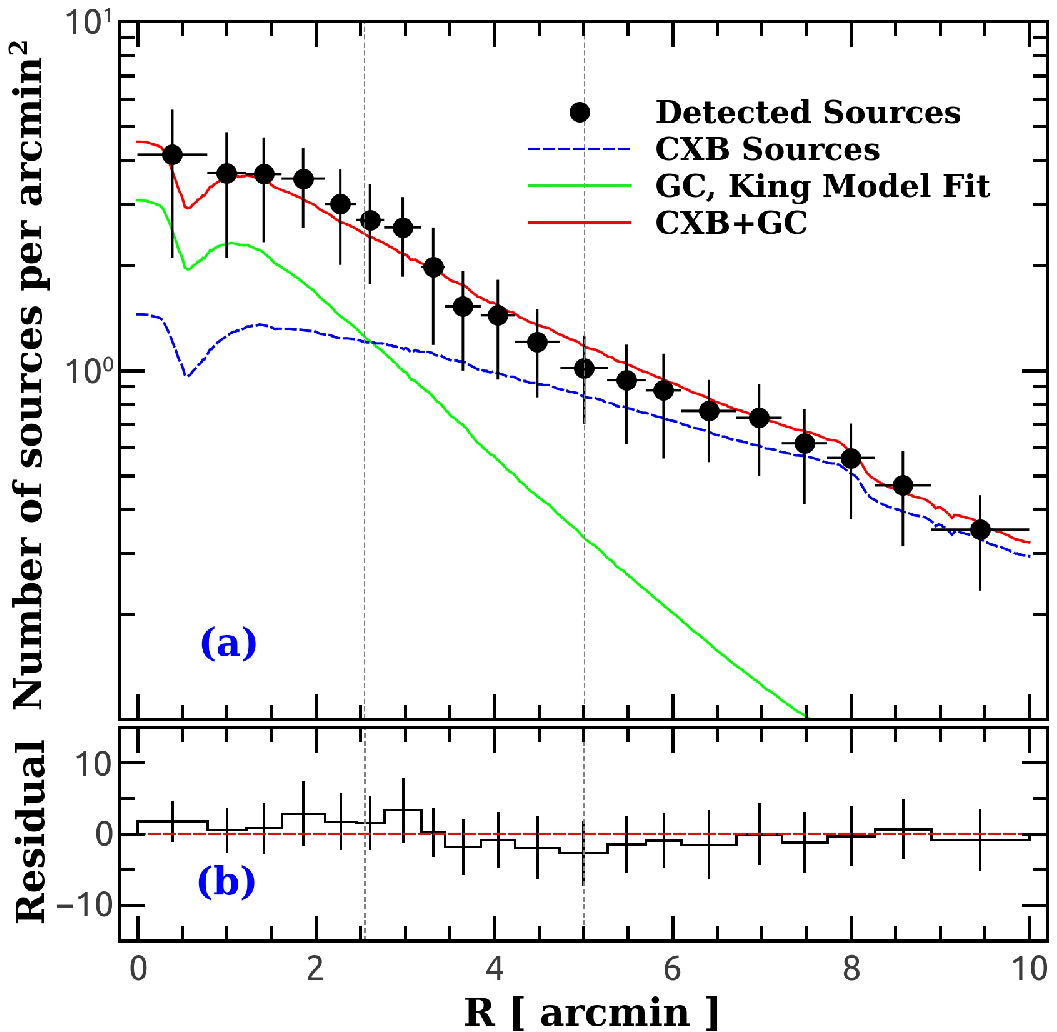}\includegraphics[angle=0,origin=br,height=0.35\textheight, width=0.51\textwidth]{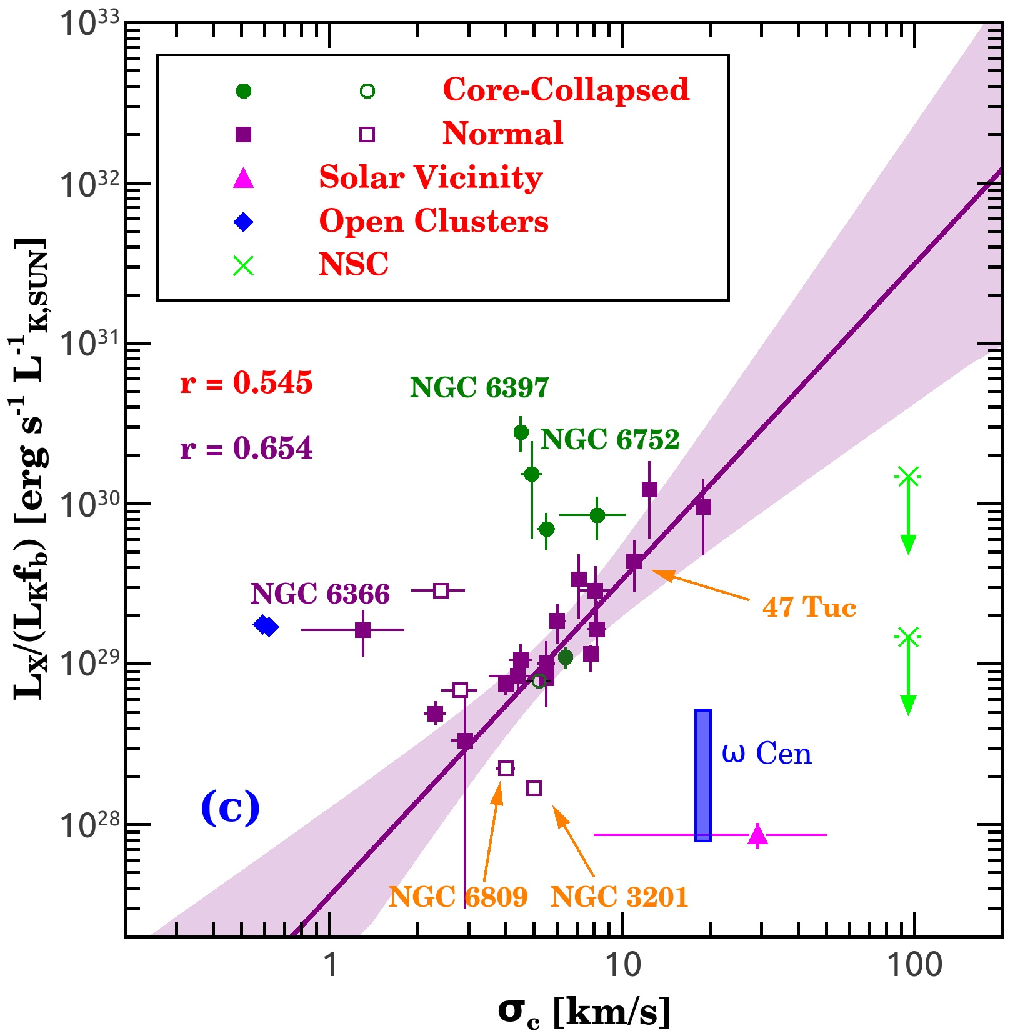}
\linespread{0.7}
\caption{(a): Radial surface density distribution of X-ray sources in $\omega$ Cen. The detected X-ray sources ($N_{\rm X}$) are plotted as black dots, while the contribution of CXB sources ($N_{\rm CXB}$) is displayed as blue dashed line. We modeled the GC component ($N_{\rm G}$; green line) with the best-fitting King model of $\omega$ Cen, which was convolved with the  limiting sensitivity function and normalized to match the number of GC X-ray sources. The red line is the sum of the CXB and GC components. (b): Residuals of $N_{\rm X}-N_{\rm CXB}-N_{\rm G}$, the horizontal red dashed line represents a uniform mixture of X-ray sources and normal stars within $\omega$ Cen. (c): Binary hardness ratio as a function of central velocity dispersion in star clusters. This image is adopted from \citet{cheng2018b}, $\omega$ Cen is marked with the blue rectangle. \label{fig:2}}
\end{figure}

From Figure-\ref{fig:1}, it can be seen that the X-ray sources are uniformly dispersed in the {\it Chandra} FoV of $\omega$ Cen, and few bright sources ($F_{p}\gtrsim 1.0\times 10^{-6}\, \rm photon\, cm^{-2} \, s^{-1}$) are located within the cluster core. These features are different from other GCs studied in this series of work (i.e., 47 Tuc, Terzan 5 and M 28; \citealp{cheng2019a, cheng2019b, cheng2020}), where the bright X-ray sources are mainly concentrated in the cluster center and significant radial distribution dips, a signature of mass segregation, can be observed in the $F_{p}-R$ diagram.
To explore the mass segregation of X-ray sources in $\omega$ Cen, we study the radial surface density profile of X-ray sources in Figure-\ref{fig:2} (a). 
Following the method used in previous work, we adopted several components to fit the observed data (black dots) in $\omega$ Cen. The first is the cosmic X-ray background (CXB) source, which is assumed to have a uniform spatial distribution and is influenced by the limiting sensitivity map of $\omega$ Cen. 
For the 295 X-ray sources located within $R<10\arcmin$, statistically 197 of them are CXB sources, estimated with the log$N$--log$S$ relation of \citet{kim2007}; while the remaining 98 sources can be reasonably accounted for by the GC X-ray sources. Since $\omega$ Cen is far away from the Galactic disk ($b\approx 15\arcdeg$), and its surface brightness (within $R<10\arcmin$) is more than 3 orders of magnitude larger than the Galactic field stars \citep{deboer2019}, here we have neglected the contribution of Galactic foreground/background sources during the fitting\footnote{\citet{haggard2009} and \citet{henleywillis2018} found some of the X-ray sources could be Galactic active stars. However, as illustrated below, the introduction of a uniformly distributed Galactic component will not change our conclusions.}.

The CXB component ($N_{\rm CXB}$) is shown as a blue dashed line in Figure-\ref{fig:2} (a). For the GC component ($N_{\rm G}$), we modeled their radial distribution with the best-fitting King model ($R_{c}=153\arcsec$, $c=1.31$; \citealp{ferraro2006}) of $\omega$ Cen, which was convolved with the limiting-sensitivity function and normalized to match the number of GC X-ray sources. The GC component is displayed as a green line, and the sum of CXB and GC components is shown as a red line in Figure-\ref{fig:2} (a). It is evident that the radial distribution of observed X-ray sources ($N_{\rm X}$) matches well with the model predicted profile. We calculate their residuals ($N_{\rm X}-N_{\rm CXB}-N_{\rm G}$) and plot it as a function of $R$ in Figure-\ref{fig:2} (b). The flat distribution of the residuals around zero suggests that the X-ray sources are uniformly mixed with MS stars in $\omega$ Cen, and there is no mass segregation of X-ray sources in this cluster. 
Our conclusions are consistent with the study of \citet{ferraro2006}, where the specific frequency of BSS is also found to be flatly distributed in $\omega$ Cen (Figure-\ref{fig:3} (b)). We leave the implication of these findings to be addressed in Section 4.1.

As suggested by \citet{ferraro2006}, the uniform mixture of BSS (X-ray sources) with normal stars in $\omega$ Cen rules out the dynamical origin of BSS (X-ray sources), and primordial binary evolution is responsible for the formation of these objects in $\omega$ Cen. 
However, using the X-ray emissivity as a proxy of the X-ray source abundance, we found a significant paucity of X-ray sources in $\omega$ Cen relative to other Galactic GCs and the galaxy field environment, represented by the Solar Neighborhood and Local Group dwarf elliptical galaxies \citep{cheng2018a}. Besides, the direct source-counting method also shows that the abundance (specific frequency) of CVs (BSS) in $\omega$ Cen is at least 2-3 times (40 times) lower than that in the Galactic field \citep{ferraro2006,haggard2009,henleywillis2018}. These evidence indicate that, even the primordial binary channel, is highly suppressed in $\omega$ Cen. According to the Hills-Heggie law, the dynamical evolution of binaries depends on the average stellar kinetic energy ($E_{k}\propto \sigma^{2}$) of the cluster, and there exists a watershed orbital separation (i.e., $a_{w}$, with $E_{b}=GM_{\ast}^{2}/2a_{w}\sim E_{k}$, and hence $a_{w}\propto\sigma^{-2}$) for MS binaries in GCs. Stellar encounters involving hard binaries ($a<a_{w}$) tend to make them harder, whereas encounters involving soft binaries ($a>a_{w}$) drive them softer and they are eventually disrupted \citep{hills1975,heggie1975}. 
If most primordial binaries were not transformed into X-ray-emitting close binaries, they may have been dynamically destroyed and served as a ``cooling mechanism" in $\omega$ Cen. Therefore, it remains an open question whether there are binary burning processes in $\omega$ Cen.

To check the binary burning processes of GCs, we have tested the Hills-Heggie law against the binary hardness ratio, defined as the relative abundance ratio of X-ray-emitting hard binaries to MS binaries and traced by $L_{\rm X}/(L_{\rm K}f_{b})$, in dozens of GCs \citep{cheng2018b}. As demonstrated in \citet{cheng2018b}, the stellar encounters of GCs are dominated by MS binaries, thus the binary hardness ratio, which could also  be regarded as the transformation rate of MS binaries into X-ray-emitting close binaries, is a sensitive probe of binary burning processes in GCs. 
With more than 15 distinct stellar populations having been identified \citep{bellini2017b}, it is hard to estimate the fraction ($f_{b}$) of MS binaries precisely in $\omega$ Cen. Based on the counting of potential MS binaries on the color--magnitude diagram (CMD) of $\omega$ Cen, \citet{bellini2017b} estimated a lower limit of $f_{b}>2.7\%$ for $\omega$ Cen. While comparing the CMD of $\omega$ Cen with simulated models, \citet{sollima2007} constrained an upper limit of $f_{b}<15\%$ for $\omega$ Cen. The X-ray emissivity of $\omega$ Cen was measured to be $L_{\rm X}/L_{\rm K}=1.26^{+0.14}_{-0.10}\times 10^{27}\, \rm erg\,s^{-1}\,L_{\rm K,\odot}^{-1}$ \citep{cheng2018a}, which corresponds to a binary hardness ratio of $7.73\times 10^{27}\, \rm erg\,s^{-1}\,L_{\rm K,\odot}^{-1}<L_{\rm X}/(L_{\rm K}f_{b})<5.19\times 10^{28}\, \rm erg\,s^{-1}\,L_{\rm K,\odot}^{-1}$.

In Figure-\ref{fig:2} (c), we compared the binary hardness ratio of $\omega$ Cen (blue rectangle) with other star clusters studied in \citet{cheng2018b}. According to the Hills-Heggie law, primordial binaries would be either transformed into X-ray-emitting close binaries or dynamically disrupted by stellar encounters, and the value of $L_{\rm X}/(L_{\rm K}f_{b})$ is sensitive to the watershed orbital separation $a_{w}$ of GCs. Therefore, the locations of star clusters on the $L_{\rm X}/(L_{\rm K}f_{b})-\sigma_{c}$ diagram are helpful in evaluating their binary burning levels. For dynamically normal GCs (purple squares), their best-fitting function (purple solid line) can be written as $L_{\rm X}/(L_{\rm K}f_{b})\propto\sigma_{c}^{1.71\pm0.48}$. The fitting slope is roughly consistent with the prediction (i.e., $a_{w}\propto \sigma^{-2}$) of the Hills-Heggie law and suggests that binary burning processes are taking place in these clusters; Core-collapsed GCs (olive dots) have a larger binary hardness ratio than the dynamically normal GCs, they are running out of their primordial binaries, and the contraction of cluster core will shorten the timescale of binary encounters, thereby boosting the extraction of energy from harder binaries; Other star clusters, such as NGC 6366 and open clusters (blue diamonds), are suffering strong tidal stripping effects and tend to disperse quickly, thus evaporation and mass segregation effects will leave these systems a large fraction of binaries and a lower stellar velocity dispersion, which drives them to evolve toward the upper left in the $L_{\rm X}/(L_{\rm K}f_{b})-\sigma_{c}$ diagram.    
Obviously, it can be seen from Figure-\ref{fig:2} (c) that $\omega$ Cen is located far below the fitting line of the dynamically normal GCs, which is opposite to the core-collapsed GCs and the open clusters, indicating that the binary burning processes are highly suppressed in this cluster. The implication of this finding will be discussed in Section 4.2.

\section{Discussion: The Role of BHS in $\omega$ Cen}
\subsection{Quenching of Mass Segregation by BHS in $\omega$ Cen}

By and large, GCs are driven to reach equipartition states by two-body relaxation, in which the velocity dispersion scales with stellar mass ($m$) as $\sigma\propto m^{-\eta}$, with $0\leq \eta\leq 0.5$ to quantify the degree of equipartition (0 for none and 0.5 for full). With long enough time, one may expect clusters to reach full equipartition. However, using a suite of direct N-body simulations, \citet{trenti2013} found clusters will never reach full equipartition, and there is a maximum value of $\eta_{max}\approx 0.15$ for the MS in GCs, while stellar remnants may reach a larger $\eta$ than $\eta_{max}$, but its value is still smaller than 0.5 in all simulations \citep{trenti2013}. The reason may lie in the effect of the gravitational potential of GCs, where slow-moving massive stars tend to fall into cluster centers, and fast-moving low-mass stars tend to migrate outward and escape from the cluster; hence only partial energy equipartition can be established in GCs. Through Monte-Carlo cluster simulations, \citet{bianchini2016} studied the mass-dependent stellar kinematics in GCs. They showing that $\sigma(m)$ can be described by an exponential function, $\sigma^{2}\propto {\rm exp}(m/m_{eq})$, with the scaling parameter $m_{eq}$ quantifying the degree of energy equipartition. In this regard, stars more massive than $m_{eq}$ are in complete equipartition ($\eta=0.5$), while stars less massive than $m_{eq}$ have values of $\eta$ that vary linearly as a function of stellar mass ($\eta=m/2m_{eq}$), and there is a tight correlation between $m_{eq}$ and the number of relaxation times ($n_{rel}$) a cluster has experienced \citep{bianchini2016}:
\begin{equation}
m_{eq}=(1.55\pm 0.23)+(4.1\pm 0.31)n_{rel}^{-0.85\pm 0.12},
\end{equation}
where $n_{rel}=T_{age}/t_{rc}$ and $T_{age}$ and $t_{rc}$ are the cluster age and core-relaxation timescale, respectively. 

Observationally, one can evaluate the dynamical evolution states of GCs either by measuring the slope $\eta$ of the $\sigma-m$ relation directly, or by studying the mass segregation of stars within GCs. 
Indeed, using the BSS as sensitive probes of mass segregation, \citet{ferraro2012} show that GCs can be classified into three families with increasing dynamical ages. In this scenario, the radial specific frequency of BSS is flatly distributed in dynamically young (Family I) GCs, and two-body relaxation will drive BSS segregation to cluster center, modifying the flat BSS distribution into a bimodal shape, with a central peak, a dip, and an outer rising branch in intermediate dynamical age (Family II) GCs. As the cluster evolves, the radial distribution dip will propagate outward gradually, and eventually leading to a BSS distribution with only a central peak in dynamically old (Family III) GCs. 

\begin{figure}[htb!]
\centering
\includegraphics[angle=0,origin=br,height=0.35\textheight, width=0.51\textwidth]{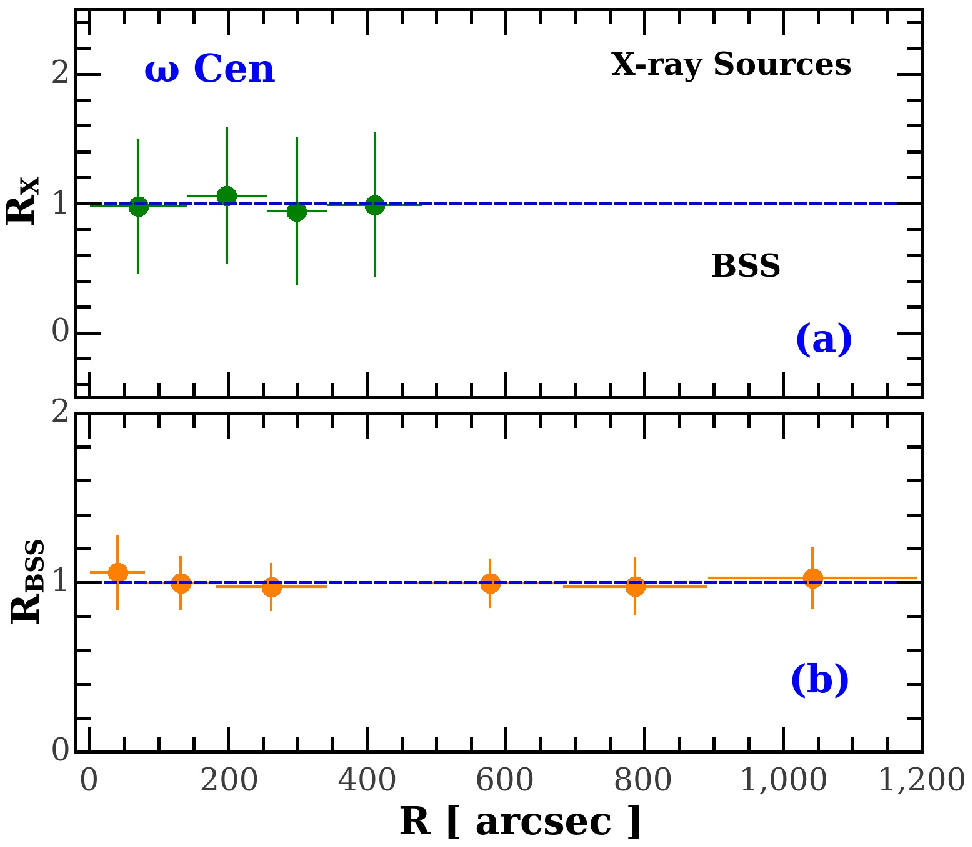}\includegraphics[angle=0,origin=br,height=0.35\textheight, width=0.51\textwidth]{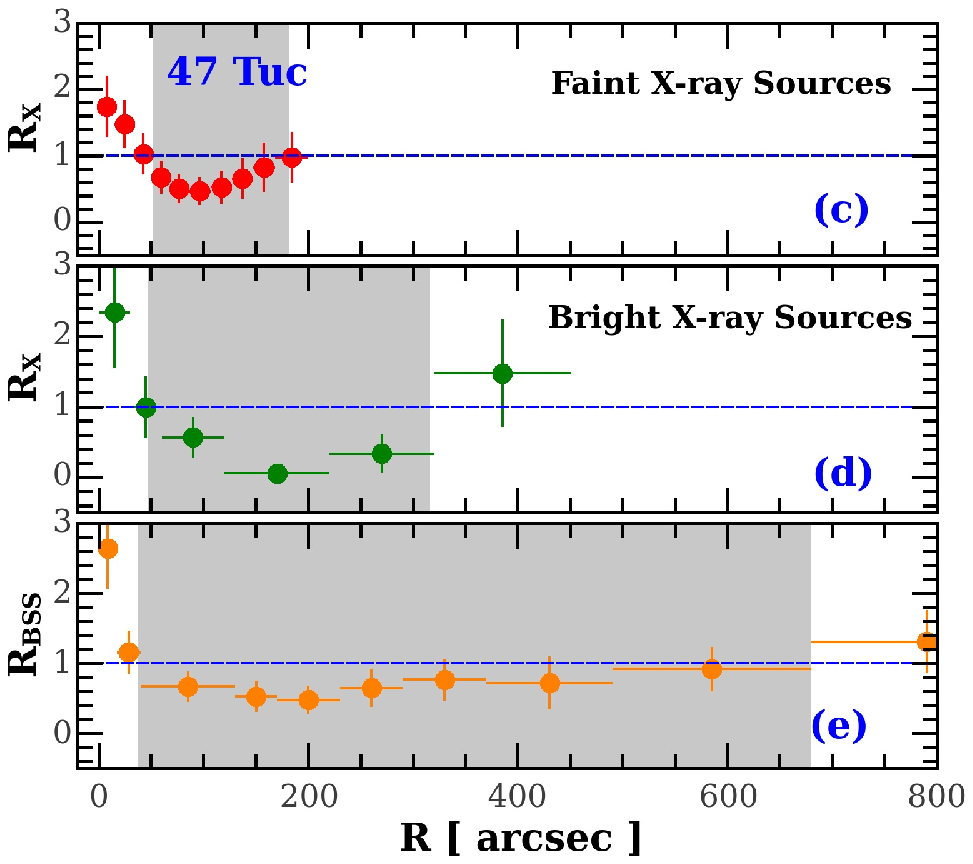}
\linespread{0.7}
\caption{A comparison of the radial specific frequency of X-ray sources and BSS in $\omega$ Cen (left panels) and 47 Tuc (right panels). For the 132 X-ray sources selected in $\omega$ Cen (i.e., olive dashed box in Figure-\ref{fig:1}), about 95 of them are CXB sources. Assuming that the CXB sources are uniformly distributed in the {\it Chandra} FoV, we corrected the CXB contamination and calculated the value of $R_{\rm X}$ using the method presented in \citet{cheng2019a}. In all panels, the blue horizontal dashed lines (with $R_{\rm X,BSS}=1$) represent a uniform distribution of X-ray sources (or BSS) relative to the integrated light of the cluster. BSS data of $\omega$ Cen were adopted from \citep{ferraro2006}, while BSS and X-ray source data of 47 Tuc were adopted from \citet{ferraro2004} and \citep{cheng2019a}, respectively. Ranges of the distribution dips (i.e. radial bins with $R_{\rm X,BSS}<1$) in 47 Tuc are highlighted with gray color. \label{fig:3}}
\end{figure}

Following the method used in \citet{cheng2019a}, we calculate the specific frequency of X-ray source in $\omega$ Cen and plot it as a function of $R$ in Figure-\ref{fig:3} (a). For comparison, we also display the radial distribution of BSS of $\omega$ Cen (with data adopted from \citealp{ferraro2006}) with orange dots in Figure-\ref{fig:3} (b). The flat radial distribution of X-ray sources and BSS suggest that $\omega$ Cen is a typical Family I cluster, which is dynamically young and that its core has not been relaxed yet \citep{ferraro2012}. 
However, using the HST/ACS observation data with a 4-year baseline, \citet{anderson2010} derived a proper-motion catalog of stars within the central region ($R<7\arcmin$) of $\omega$ Cen. They found a value of $\eta\approx 0.2$ for stars with mass of $0.5\,M_{\odot}\leq m\leq 0.8\, M_{\odot}$ in $\omega$ Cen. While extending the time-baseline to 8 years, \citet{bellini2013} refined the proper-motion catalogue of \citet{anderson2010} and derived a value of $\eta=0.16\pm 0.05$ for stars in the mass range of $0.3\,M_{\odot}\leq m\leq 0.8\, M_{\odot}$ \citep{bellini2018}. 
On the other hand, by comparing the proper-motion of BSS and MS turnoff (MSTO) stars directly, \citet{baldwin2016} show that the velocity dispersion of BSS is comparable to that of the MSTO stars in $\omega$ Cen, with $\sigma_{\rm BSS}/\sigma_{\rm MSTO}=1.02\pm0.04$. 
These evidence suggest that MS stars have reached their peak value, $\eta_{max}$, at the central region of $\omega$ Cen \citep{trenti2013}, but the mass segregation of BSS (X-ray sources) is inhibited in this cluster. 

A possible explanation is that there is a BHS in $\omega$ Cen. With a much larger average mass than BSS and X-ray sources, the BHs are expected to settle in the cluster center quickly and form a BHS, which may influence the mass segregation of BSS and X-ray sources in two ways. First, the BHS may serve as a dominating energy source in GCs, and its thermal coupling with the host cluster may drive the cluster to expand \citep{mackey2008, giersz2019}. Thus GCs with a BHS are more likely to have a puffy structure, which enhances the two-body relaxation timescale and slows down the mass segregation of BSS and X-ray sources in GCs \citep{alessandrini2016, peuten2016, weatherford2018, arcasedda2018}. Second, the BHs may quench the mass segregation of heavy stars (such as BSS, X-ray sources and MS binaries) by directly expelling them out of the cluster center via gravitational encounters, so that energy can be effectively transferred from the BHS to the host cluster and launch the outflow of energy in GCs. As a result, it is the heaviest objects ($m\sim m_{eq}$) that determine the stellar equipartition and dominate the mass segregation ($d\sigma/\sigma\propto dm/m_{eq}$) of GCs \citep{bianchini2016}. 

To illustrate this effect better, we make a comparison between $\omega$ Cen and 47 Tuc in Figure-\ref{fig:3}. 47 Tuc is a typical Family II cluster, showing significant distribution dips of BSS and X-ray sources identified by \citet{ferraro2004} and \citet{cheng2019a} respectively. The proper-motion study shows that 47 Tuc has reached a partial equipartition state. With $\eta$ is measured to be $\eta=0.21\pm0.01$ in the mass range of $0.3\,M_{\odot}\leq m\leq 0.9\, M_{\odot}$ \citep{watkins2020}. According to the value of $\eta$, the evolution state of 47 Tuc is comparable to $\omega$ Cen. However, the radial distribution of BSS and X-ray sources suggests that these two clusters are apparently different. With an age of $T_{age}=13.06\, \rm Gyr$ \citep{forbes2010} and a core relaxation timescale of $t_{rc}\approx 0.69\, \rm Gyr$ \citep{harris1996}, 47 Tuc is dynamically older ($n_{rel}=190$) and its scaling parameter is estimated to be $m_{eq}\approx 1.6\, M_{\odot}$. The lower value of $m_{eq}$ indicates that 47 Tuc has exhausted its BH population, and low-mass stars (such as BSS and X-ray sources) started to fall into the cluster core and created significant distribution dips in this cluster. Indeed, the proper-motion study shows that BSS are successfully separated from the MSTO stars in 47 Tuc, with $\sigma_{\rm BSS}/\sigma_{\rm MSTO}=0.77\pm0.06$ \citep{baldwin2016}.
More interestingly, we found that there is evident mass segregation delay between the BSS, bright and faint X-ray sources in 47 Tuc \citep{cheng2019a}. Although the BSS ($m=1.51\pm0.17\, M_{\odot}$) have a larger average mass than the bright ($m=1.44\pm0.15\, M_{\odot}$) and the faint ($m=1.16\pm0.06\, M_{\odot}$) X-ray sources. It seems that BSS can not quench the mass segregation of the X-ray sources effectively. With $n_{rel}>>1$, X-ray sources had been expelled out of the cluster core would be quickly dragged back to the cluster center by dynamical friction, and created a distribution dip within the region avoids of BSS in 47 Tuc (Figure-\ref{fig:3} (c), (d) and (e)). 
For $\omega$ Cen, the core relaxation timescale ($t_{rc}\approx 3.98\, \rm Gyr$; \citealp{harris1996}) is about 2.9 times shorter than its age ($T_{age}\approx 11.52\, \rm Gyr$; \citealp{forbes2010}), which corresponds to a scaling parameter of $m_{eq} \approx3.2\, M_{\odot}$. 
The large value of $m_{eq}$ is in support of the presence of a BHS (with $m_{\bullet}\sim 3.2\, M_{\odot}$) in $\omega$ Cen\footnote{Here the adopted BH's average mass is a rough estimate, which is biased toward the low-mass end of the BH mass function.}. Moreover, as the value of $n_{rel}$ decreases quickly outside of the cluster core, dynamical friction is inefficient in driving BSS and X-ray sources fall back to the cluster center when they are expelled out of the cluster core by gravitational encounters. Therefore, both BSS and X-ray sources are uniformly mixed with normal stars in $\omega$ Cen.

\subsection{Black Hole Burning as an Interanl Energy Source in $\omega$ Cen}
 
Among the Galactic GCs, $\omega$ Cen is unique for many reasons. It is one of the few clusters found to have a retrograde orbit in the Milky Way \citep{dinescu1999, vasiliev2019}, and its stellar metallicity and population is also the most complicated \citep{lee1999, smith2000}, with at least 15 subpopulations have been identified \citep{bellini2017b}. These observational features have led to the suggestion that $\omega$ Cen is the remnant core of a dwarf galaxy that was captured and almost completely tidally stripped by the Milky Way \citep{majewski2000, bekki2003}. Theoretically, the Galactic tidal stripping is thought to have a net effect in accelerating the dynamical evolution of clusters \citep{gnedin1999, baumgardt2003, gieles2011}, which favors the production of energy in GCs and may be responsible for the creation of overabundant binary burning products in Terzan 5 \citep{cheng2019b}. 
However, this picture is at odds with the case of $\omega$ Cen. In Section 3, we have shown that the abundance of weak X-ray sources and BSS in $\omega$ Cen is much smaller than in the Galactic field, Local Group dwarf elliptical galaxies and other GCs, and the binary hardness ratio of $\omega$ Cen is located far below the fitting correlation line of dynamically normal GCs in the $L_{\rm X}/(L_{\rm K}f_{b})-\sigma_{c}$ diagram. These evidence suggest that the binary burning processes are highly suppressed in $\omega$ Cen, and other internal energy sources are dominating the energy production of this cluster. In fact, although the orbit of $\omega$ Cen is found to be highly eccentric (i.e., with perigalactic distance $R_{\rm p}=1.35\pm 0.04$ kpc and apogalactic distance $R_{\rm a}=7.0\pm 0.04$ kpc, respectively; \citealp{baumgardt2019b}), and the cluster tidal radius is about 3 times larger at apogalacticon than at perigalacticon. Significant tidal stream was identified to be associated with $\omega$ Cen around the apogalactic passage \citep{myeong2018,ibata2019}. These features lend support to the existence of an invisible ``heating mechanism" in $\omega$ Cen, which drives the cluster to expand to be tidally limited throughout the orbital motion in the Milky Way.

In addition to the binary burning processes, there are three other ``heating mechanisms" in GCs. The first is the strong mass-loss due to rapid stellar evolution, which heats the cluster indirectly by reducing the potential well of GCs quickly, thereby driving the system to expand in the early stage of cluster evolution \citep{mackey2008,contenta2015}. Once the massive stars have evolved, the BHs may concentrate to cluster center and form a BHS. Through BH burning processes, BHs are frequently ejected to higher orbits in the cluster potential, leading to interactions with luminous stars in the outer parts of GCs. BHs therefore can deposit energy into the host cluster and dominate the expansion of GCs in later times \citep{merritt2004, mackey2007, mackey2008, breen2013, wang2016, arcasedda2018, giersz2019}. 
Finally, if a central IMBH can be successfully formed through the repeated mergers (or accretion) of BHs and stars within GCs, simulations suggest that it can also serve as a ``heating mechanism" \citep{baumgardt2004, trenti2007,gill2008}. Since $\omega$ Cen is $\sim 11.5 \, \rm Gyr$ old and the current mass-loss is negligible, we therefore only consider the cases of BHS and IMBH below.

In fact, the cluster surface brightness and stellar velocity dispersion profile have provided strong evidence for the existence of a dark component in the center of $\omega$ Cen, which could be either a BHS or an IMBH and both can have a contribution to the gravitational mass of the system \citep{noyola2008, jalali2012, vandermarel2010, baumgardt2017, zocchi2019}. 
From the aspect of cluster dynamical evolution, it is hard to discriminate the BHS and IMBH scenarios with full certainty, since both scenarios have similar effect in heating the cluster and quenching the mass segregation of low-mass stars in GCs \citep{baumgardt2004,gill2008,trenti2013,devita2019}. However, \citet{baumgardt2019a} argued that GCs hosting an IMBH tend to have a deeper central potential well, thus more energetic stars are expected to be observed near the cluster center. In this regard, they suggested that a BHS is better than the IMBH scenario in interpreting the absence of more energetic stars (with 1D velocity $v>60 \,\rm km\,s^{-1}$) within the central 20 arcsec of $\omega$ Cen \citep{baumgardt2019a}.

Suppression of binary burning processes in $\omega$ Cen may provide another perspective to discriminate the BHS and IMBH scenarios. As discussed in \citet{cheng2018b}, the dynamical evolution of primordial binaries in GCs is subject to the Hills-Heggie law, which states that the hardening or disruption of binaries depends on the stellar velocity dispersion of the cluster. 
If there is no BHS in $\omega$ Cen, the primordial binaries are mainly disrupted or hardened via the dynamical encounters with other normal stars, and the observed binary hardness ratio of $\omega$ Cen should be larger, with the value of $L_{\rm X}/(L_{\rm K}f_{b})$ follow the correlation line predicted by the Hills-Heggie law in Figure-\ref{fig:2} (c). On the other hand, the presence of a BHS would have a net effect of reducing the value of $L_{\rm X}/(L_{\rm K}f_{b})$ in $\omega$ Cen. Through exchange encounters involving a BH, hard primordial binaries are preferentially transformed into BH-MS\footnote{Due to the low duty cycles for the active state of a mass-transferring BH \citep{kalogera2004}, BH-MS systems are hard to find via X-ray emission, and their contribution to GC X-ray source population is small.} or BH binaries, thus resulting in a much lower abundance of binary burning products in $\omega$ Cen. We note that this picture is consistent with the simulations of \citet{ye2019} and \citet{kremer2018b,kremer2020b}, where the formation efficiency of LMXBs, MSPs, CVs, ABs and BSS is found to be inhibited by the presence of a BHS in GCs. 

With an IMBH, one may find from Figure-\ref{fig:2} (c) a lot of similarity between $\omega$ Cen and the Galactic Nuclear Star Cluster (NSC), in which the central supermassive BH ($M_{\bullet}\sim 4\times 10^{6}\, \rm M_{\odot}$) enhances the stellar kinetic energy greatly, thereby boosting the dynamical disruption of binaries and reducing the observed value of $L_{\rm X}/(L_{\rm K}f_{b})$ in NSC \citep{cheng2018b}. However, the fraction of the IMBH mass ($M_{\bullet}<4.75\times 10^{4}\, \rm M_{\odot}$) in $\omega$ Cen is less than $5\%$ of the cluster mass \citep{baumgardt2019a}, and its contribution to stellar kinetic energy is small. Therefore, although an IMBH is in favor of disrupting binaries in GCs, most of the primordial binaries are destroyed by other stars rather than by the IMBH directly \citep{trenti2007}, which has a small effect in reducing the value of $L_{\rm X}/(L_{\rm K}f_{b})$ in $\omega$ Cen. Furthermore, if an IMBH is dominating the energy production of $\omega$ Cen, massive stars would fall into the radius of influence ($R_{roi}$) of the IMBH and form a density cusp \citep{baumgardt2004}. But in Figure-\ref{fig:1}, only one faint X-ray source was found to be located within the margin of $R_{roi}$ in $\omega$ Cen. 
Taking these aspects together, we suggest that the BHS is the most likely scenario of ``heating mechanism" in $\omega$ Cen. Through the BH burning processes, the BHS is dominating the energy production of $\omega$ Cen, which also makes $\omega$ Cen a promising factory of gravitational-wave sources in the Galaxy.

For comparison, we also marked the locations of NGC 3201, NGC 6809 and 47 Tuc with orange texts in Figure-\ref{fig:2} (c). NGC 3201 and NGC 6809 are found to have an upper limit of $L_{\rm X}/(L_{\rm K}f_{b})$ smaller than that predicted by the fitting correlation of dynamically normal GCs \citep{cheng2018b}, which indicates that the binary burning processes are also suppressed in these two GCs. For NGC 3201, three BH candidates are found in detached BH-MS binary systems \citep{giesers2018,giesers2019}, strongly supporting the existence of a BHS and BH burning in this cluster. 
While for 47 Tuc, it is dynamically older and the BH population has been exhausted, hence MS binaries are dominating the production of energy in this cluster, which is responsible for the creation of a large population of binary burning products in 47 Tuc \citep{cheng2019a}.

\section{Summary}

Using the X-ray sources as sensitive probes of stellar dynamical interactions in GCs, we study the mass segregation and binary burning processes in $\omega$ Cen. Our main findings are as follows:

1. Although the MS stars have reached a higher partial equipartition state ($\eta=0.16\pm0.05$) in the central region of $\omega$ Cen. We show that the mass segregation of X-ray sources is quenched in this cluster, and the radial distribution of X-ray sources is flatly distributed with respect to the normal stars in $\omega$ Cen.

2. The X-ray source abundance in $\omega$ Cen is much smaller than that in the Galactic field and other GCs \citep{cheng2018a}, and the binary hardness ratio (traced by $L_{\rm X}/(L_{\rm K}f_{b})$) of $\omega$ Cen is far below the value predicted by the Hills-Heggie law in the $L_{\rm X}/(L_{\rm K}f_{b})-\sigma_{c}$ diagram. These features suggest that the binary burning processes are highly suppressed in $\omega$ Cen, and other ``heating mechanisms", either a BHS or an IMBH, are essential in driving the dynamical evolution of $\omega$ Cen.

3. Compared with the IMBH, the BHS scenario may provide a better explanation for the observed features in $\omega$ Cen. Mass segregation of X-ray sources can be effectively quenched by the presence of a BHS in $\omega$ Cen. Meanwhile, through exchange encounters, the BHS can also suppress the formation efficiency of binary burning products and reduce the value of $L_{\rm X}/(L_{\rm K}f_{b})$ in $\omega$ Cen. We note that these features are indirect evidence of BHS in $\omega$ Cen, and future detection of BHs by microlensing can be used to constrain the BH populations in this cluster \citep{zaris2020}.

4. Through the BH burning processes, the BHS is dominating the energy production of $\omega$ Cen, which also makes $\omega$ Cen a promising factory of gravitational-wave sources in the Galaxy, and future gravitational-wave observations may provide direct evidence to verify the BH burning processes in GCs \citep{benacquista2001, kremer2018c}.

\begin{deluxetable*}{cllccccccccc}
\tabletypesize{\scriptsize}
\tablewidth{0pt}
\tablecaption{Main {\it Chandra} Source Catalog}
\tablecolumns{12}
\linespread{1}
\tablewidth{0pc}
\tablenum{1}
\tablehead{
\colhead{XID} & \colhead{RA} & \colhead{Dec} & \colhead{Error} & \colhead{R} & \colhead{$\log P_{\rm B}$} & \colhead{$C_{net,f}$} & \colhead{$C_{net,s}$} & \colhead{$F_{p,f}$} & \colhead{$F_{p,s}$} & \colhead{$\log L_{X,f}$} & \colhead{$\log L_{X,s}$} \\
\colhead{} & \colhead{(\arcdeg)} & \colhead{(\arcdeg)} & \colhead{(\arcsec)} & \colhead{(\arcsec)} & \colhead{} & \colhead{(cts)}  & \colhead{(cts)}  & \colhead{(ph/s/cm$^{2}$)} & \colhead{(ph/s/cm$^{2}$)} & \colhead{(erg/s)} & \colhead{(erg/s)} \\
\colhead{(1)} & \colhead{(2)} & \colhead{(3)} & \colhead{(4)} &\colhead{(5)} & \colhead{(6)} & \colhead{(7)} & \colhead{(8)} & \colhead{(9)} & \colhead{(10)} & \colhead{(11)} & \colhead{(12)}
}
\startdata
1 & 201.454043 & -47.524603 & 0.7 & 612.3 & $<$-5 & $56.9^{+8.6}_{-8.5}$   & $49.3^{+7.7}_{-7.3}$   & 4.25E-6 & 2.45E-6 & 31.59 & 31.53 \\ % (1)
2 & 201.488797 & -47.547074 & 0.3 & 561.2 & $<$-5 & $184^{+15}_{-14}$      & $133^{+12}_{-12}$      & 7.16E-6 & 3.90E-6 & 31.90 & 31.65 \\ % (2)
3 & 201.497022 & -47.563423 & 0.6 & 571.9 & $<$-5 & $38.5^{+9.8}_{-9.3}$   & $21.6^{+6.0}_{-5.6}$   & 8.71E-7 & 4.09E-7 & 31.05 & 30.53 \\ % (3)
4 & 201.502950 & -47.486434 & 0.4 & 472.3 & $<$-5 & $58.1^{+10.1}_{-10.0}$ & $39.1^{+8.2}_{-7.6}$   & 1.10E-6 & 6.02E-7 & 31.01 & 30.84 \\ % (4)
5 & 201.506623 & -47.551621 & 0.1 & 530.3 & $<$-5 & $2138^{+52}_{-48}$     & $1054^{+36}_{-34}$     & 3.99E-5 & 1.54E-5 & 32.90 & 32.60 \\ % (5)
6 & 201.516903 & -47.587070 & 0.7 & 584.0 & $<$-5 & $35.2^{+9.8}_{-9.3}$   & $21.5^{+6.3}_{-5.7}$   & 8.51E-7 & 4.29E-7 & 31.03 & 30.51 \\ % (6)
7 & 201.517651 & -47.468022 & 0.5 & 438.0 & -4.66 & $27.3^{+8.4}_{-7.8}$   & $18.0^{+6.2}_{-5.5}$   & 4.76E-7 & 2.48E-7 & 30.78 & 30.26 \\ % (7)
8 & 201.518990 & -47.461417 & 0.2 & 437.7 & $<$-5 & $149^{+14}_{-13}$      & $97.2^{+10.8}_{-10.7}$ & 2.60E-6 & 1.33E-6 & 31.46 & 31.08 \\ % (8)
\enddata
%\vspace{-0.5cm}
\tablecomments{
Column\ (1): sequence number of the X-ray source catalog, sorted by R.A.
Columns.\ (2) and (3): right ascension and decl. for epoch J2000.0.
Columns.\ (4) and (5): estimated standard deviation of the source position error and its projected distance from cluster center.
Column.\ (6): logarithmic Poisson probability of a detection not being a source.
Columns.\ (7) -- (10): net source counts and photon flux extracted in the full (0.5--8~keV) and soft (0.5--2~keV) band, respectively.
Columns.\ (11) and (12): unabsorbed source luminosity in full and soft band. 
}
\label{tab:mcat}
\end{deluxetable*}

\begin{acknowledgements}
We thank the anonymous referee for the valuable comments that helped to improve our manuscript. This work is supported by the Youth Program of National Natural Science Foundation of China No. 12003017, the Fundamental Research Funds for the Central Universities No. 2042020kf0035, the National Key Research and Development Program of China No. 2016YFA0400803, the Natural Science Foundation of China under grants No. 11773015, 11622326, and the Project U1838201, U1838103 supported by NSFC and CAS. 

\end{acknowledgements}

%\bibliographystyle{aasjournal} 

%\newpage
%\clearpage
\label{lastpage}


\begin{thebibliography}{dummy}
%\expandafter\ifx\csname natexlab\endcsname\relax\def\natexlab#1{#1}\fi

\bibitem[Anderson \& van der Marel(2010)]{anderson2010} Anderson, J., \& van der Marel, R.~P.\ 2010, \apj, 710, 1032
\bibitem[Alessandrini et al.(2016)]{alessandrini2016} Alessandrini, E., Lanzoni, B., Ferraro, F.~R., Miocchi, P., \& Vesperini, E.\ 2016, \apj, 833, 252 
\bibitem[Antonini et al.(2019)]{antonini2019} Antonini, F., Gieles, M., \& Gualandris, A.\ 2019, \mnras, 486, 5008
\bibitem[Antonini \& Gieles(2020)]{antonini2020} Antonini, F., \& Gieles, M.\ 2020, \mnras, 492, 2936
\bibitem[Arca Sedda et al.(2018)]{arcasedda2018} Arca Sedda, M., Askar, A., \& Giersz, M.\ 2018, \mnras, 479, 4652
\bibitem[Askar et al.(2017)]{askar2017} Askar, A., Szkudlarek, M., Gondek-Rosi{\'n}ska, D., Giersz, M., \& Bulik, T.\ 2017, \mnras, 464, L36 
\bibitem[Askar et al.(2018)]{askar2018} Askar, A., Arca Sedda, M., \& Giersz, M.\ 2018, \mnras, 478, 1844
\bibitem[Askar et al.(2019)]{askar2019} Askar, A., Askar, A., Pasquato, M., et al.\ 2019, \mnras, 485, 5345
\bibitem[Askar et al.(2020)]{askar2020} Askar, A., Giersz, M., Arca Sedda, M., et al.\ 2020, in IAU Symposium. 351, Star Clusters: From the Milky Way to the Early Universe, ed. A. Bragaglia et al. (Cambridge: Cambridge Univ. Press), 395
\bibitem[Bahramian et al.(2017)]{bahramian2017} Bahramian, A., Heinke, C.~O., Tudor, V., et al.\ 2017, \mnras, 467, 2199
\bibitem[Baldwin et al.(2016)]{baldwin2016} Baldwin, A.~T., Watkins, L.~L., van der Marel, R.~P., et al.\ 2016, \apj, 827, 12
\bibitem[Banerjee et al.(2010)]{banerjee2010} Banerjee, S., Baumgardt, H., \& Kroupa, P.\ 2010, \mnras, 402, 371
\bibitem[Baumgardt \& Hilker(2018)]{baumgardt2018} Baumgardt, H., \& Hilker, M.\ 2018, \mnras, 478, 1520
\bibitem[Baumgardt \& Makino(2003)]{baumgardt2003} Baumgardt, H. \& Makino, J.\ 2003, \mnras, 340, 227
\bibitem[Baumgardt et al.(2004)]{baumgardt2004} Baumgardt, H., Makino, J., \& Ebisuzaki, T.\ 2004, \apj, 613, 1143
\bibitem[Baumgardt(2017)]{baumgardt2017} Baumgardt, H.\ 2017, \mnras, 464, 2174
\bibitem[Baumgardt et al.(2019a)]{baumgardt2019a} Baumgardt, H., He, C., Sweet, S.~M., et al.\ 2019a, \mnras, 488, 5340 
\bibitem[Baumgardt et al.(2019b)]{baumgardt2019b} Baumgardt, H., Hilker, M., Sollima, A., \& Bellini, A.\ 2019b, \mnras, 482, 5138 
\bibitem[Bekki \& Freeman(2003)]{bekki2003} Bekki, K., \& Freeman, K.~C.\ 2003, \mnras, 346, L11
\bibitem[Bellini et al.(2013)]{bellini2013} Bellini, A., van der Marel, R.~P., \& Anderson, J.\ 2013, \memsai, 84, 140
\bibitem[Bellini et al.(2017a)]{bellini2017a} Bellini, A., Anderson, J., Bedin, L.~R., et al.\ 2017a, \apj, 842, 6
\bibitem[Bellini et al.(2017b)]{bellini2017b} Bellini, A., Milone, A.~P., Anderson, J., et al.\ 2017b, \apj, 844, 164
\bibitem[Bellini et al.(2018)]{bellini2018} Bellini, A., Libralato, M., Bedin, L.~R., et al.\ 2018, \apj, 853, 86
\bibitem[Belloni et al.(2016)]{belloni2016} Belloni, D., Giersz, M., Askar, A., et al.\ 2016, \mnras, 462, 2950
\bibitem[Belloni et al.(2017)]{belloni2017} Belloni, D., Zorotovic, M., Schreiber, M.~R., et al.\ 2017, \mnras, 468, 2429
\bibitem[Belloni et al.(2019)]{belloni2019} Belloni, D., Giersz, M., Rivera Sandoval, L.~E., et al.\ 2019, \mnras, 483, 315
\bibitem[Benacquista et al.(2001)]{benacquista2001} Benacquista, M.~J., Portegies Zwart, S., \& Rasio, F.~A.\ 2001, Classical and Quantum Gravity, 18, 4025
\bibitem[Bianchini et al.(2016)]{bianchini2016} Bianchini, P., van de Ven, G., Norris, M.~A., et al.\ 2016, \mnras, 458, 3644 
\bibitem[Breen, \& Heggie(2013)]{breen2013} Breen, P.~G., \& Heggie, D.~C.\ 2013, \mnras, 432, 2779
\bibitem[Broos et al.(2010)]{broos2010} Broos P. S., Townsley L. K., Feigelson E. D., Getman K. V., Bauer F. E., Garmire G. P., 2010, \apj, 714, 1582
\bibitem[Camilo \& Rasio(2005)]{camilo2005} Camilo, F., \& Rasio, F. A. 2005, in ASP Conf. Ser. 328, Binary Radio Pulsars (San Francisco, CA: ASP), 147
\bibitem[Chatterjee et al.(2013)]{chatterjee2013} Chatterjee, S., Rasio, F.~A., Sills, A., et al.\ 2013, \apj, 777, 106
\bibitem[Chatterjee et al.(2017)]{chatterjee2017} Chatterjee, S., Rodriguez, C.~L., \& Rasio, F.~A.\ 2017, \apj, 834, 68
\bibitem[Cheng et al.(2018a)]{cheng2018a} Cheng, Z., Li, Z., Xu, X., \& Li, X.\ 2018a, \apj, 858, 33 
\bibitem[Cheng et al.(2018b)]{cheng2018b} Cheng, Z., Li, Z., Xu, X., et al.\ 2018b, \apj, 869, 52 
\bibitem[Cheng et al.(2019a)]{cheng2019a} Cheng, Z., Li, Z., Li, X., Xu, X., \& Fang, T.\ 2019a, \apj, 876, 59 
\bibitem[Cheng et al.(2019b)]{cheng2019b} Cheng, Z., Li, Z., Fang, T., et al.\ 2019b, \apj, 883, 90
\bibitem[Cheng et al.(2020)]{cheng2020} Cheng, Z., Mu, H., Li, Z., et al.\ 2020, \apj, 892, 16
\bibitem[Chomiuk et al.(2013)]{chomiuk2013} Chomiuk, L., Strader, J., Maccarone, T.~J., et al.\ 2013, \apj, 777, 69
\bibitem[Clark(1975)]{clark1975} Clark, G. W. 1975, \apj, 199, L143
\bibitem[Contenta et al.(2015)]{contenta2015} Contenta, F., Varri, A.~L., \& Heggie, D.~C.\ 2015, \mnras, 449, L100
\bibitem[de Boer et al.(2019)]{deboer2019} de Boer, T.~J.~L., Gieles, M., Balbinot, E., et al.\ 2019, \mnras, 485, 4906
\bibitem[de Vita et al.(2019)]{devita2019} de Vita, R., Trenti, M., \& MacLeod, M.\ 2019, \mnras, 485, 5752
\bibitem[Downing et al.(2010)]{downing2010} Downing, J.~M.~B., Benacquista, M.~J., Giersz, M., et al.\ 2010, \mnras, 407, 1946
\bibitem[Dinescu et al.(1999)]{dinescu1999} Dinescu, D.~I., van Altena, W.~F., Girard, T.~M., et al.\ 1999, \aj, 117, 277

\bibitem[Eddington(1926)]{eddington1926} Eddington, A.~S.\ 1926, The Internal Constitution of the Stars
\bibitem[Ferraro et al.(2004)]{ferraro2004} Ferraro, F. R., Beccari, G., et al. 2004, \apj, 603, 127
\bibitem[Ferraro et al.(2006)]{ferraro2006} Ferraro, F.~R., Sollima, A., Rood, R.~T., et al.\ 2006, \apj, 638, 433 
\bibitem[Ferraro et al.(2012)]{ferraro2012} Ferraro, F. R., Lanzoni, B., Dalessandro, E., et al. 2012, \nat, 492, 393
\bibitem[Forbes \& Bridges(2010)]{forbes2010} Forbes, D.~A., \& Bridges, T.\ 2010, \mnras, 404, 1203 
\bibitem[Fregeau et al.(2003)]{fregeau2003} Fregeau, J.~M., G{\"u}rkan, M.~A., Joshi, K.~J., \& Rasio, F.~A.\ 2003, \apj, 593, 772 
\bibitem[Fregeau et al.(2004)]{fregeau2004} Fregeau, J.~M., Cheung, P., Portegies Zwart, S.~F., \& Rasio, F.~A.\ 2004, \mnras, 352, 1 
\bibitem[Gieles et al.(2011)]{gieles2011} Gieles, M., Heggie, D.~C., \& Zhao, H.\ 2011, \mnras, 413, 2509
\bibitem[Giersz et al.(2015)]{giersz2015} Giersz, M., Leigh, N., Hypki, A., et al.\ 2015, \mnras, 454, 3150
\bibitem[Giersz et al.(2019)]{giersz2019} Giersz, M., Askar, A., Wang, L., et al.\ 2019, \mnras, 487, 2412
\bibitem[Giesers et al.(2018)]{giesers2018} Giesers, B., Dreizler, S., Husser, T.-O., et al.\ 2018, \mnras, 475, L15
\bibitem[Giesers et al.(2019)]{giesers2019} Giesers, B., Kamann, S., Dreizler, S., et al.\ 2019, \aap, 632, A3
\bibitem[Giesler et al.(2018)]{giesler2018} Giesler, M., Clausen, D., \& Ott, C.~D.\ 2018, \mnras, 477, 1853
\bibitem[Gill et al.(2008)]{gill2008} Gill, M., Trenti, M., Miller, M.~C., et al.\ 2008, \apj, 686, 303
\bibitem[Gnedin et al.(1999)]{gnedin1999} Gnedin, O.~Y., Lee, H.~M., \& Ostriker, J.~P.\ 1999, \apj, 522, 935 
\bibitem[Haggard et al.(2009)]{haggard2009} Haggard, D., Cool, A.~M., \& Davies, M.~B.\ 2009, \apj, 697, 224
\bibitem[Harris(2010 edition)]{harris1996} Harris, W. E. 1996(2010 edition), \aj, 112, 1487
\bibitem[Henleywillis et al.(2018)]{henleywillis2018} Henleywillis, S., Cool, A.~M., Haggard, D., et al.\ 2018, \mnras, 479, 2834
\bibitem[H{\'e}non(1975)]{henon1975} H{\'e}non, M.\ 1975, Dynamics of the Solar Systems, 133
\bibitem[Heggie(1975)]{heggie1975} Heggie, D. C., 1975, \mnras, 173, 729
\bibitem[Heggie \& Hut(2003)]{heggie2003} Heggie D. C., Hut P. 2003, The Gravitational Million-Body Problem: A Multidisciplinary Approach to Star Cluster Dynamics (Cambridge: Cambridge University Press)
\bibitem[Heinke et al.(2020)]{heinke2020} Heinke, C.~O., Ivanov, M.~G., Koch, E.~W., et al.\ 2020, \mnras, 492, 5684
\bibitem[Hills(1975)]{hills1975} Hills, J. G., 1975, \aj, 80, 809
\bibitem[Hurley(2007)]{hurley2007} Hurley, J.~R.\ 2007, \mnras, 379, 93
\bibitem[Hong et al.(2017)]{hong2017} Hong, J., Vesperini, E., Belloni, D., \& Giersz, M.\ 2017, \mnras, 464, 2511 
\bibitem[Hong et al.(2018)]{hong2018} Hong, J., Vesperini, E., Askar, A., et al.\ 2018, \mnras, 480, 5645
\bibitem[Ibata et al.(2019)]{ibata2019} Ibata, R.~A., Bellazzini, M., Malhan, K., et al.\ 2019, Nature Astronomy, 3, 667
\bibitem[Ivanova et al.(2006)]{ivanova2006} Ivanova, N., Heinke, C.~O., Rasio, F.~A., et al.\ 2006, \mnras, 372, 1043 
\bibitem[Ivanova et al.(2008)]{ivanova2008} Ivanova, N., Heinke, C.~O., Rasio, F.~A., Belczynski, K., \& Fregeau, J.~M.\ 2008, \mnras, 386, 553 
\bibitem[Ivanova et al.(2010)]{ivanova2010} Ivanova, N., Chaichenets, S., Fregeau, J., et al.\ 2010, \apj, 717, 948 
\bibitem[Jalali et al.(2012)]{jalali2012} Jalali, B., Baumgardt, H., Kissler-Patig, M., et al.\ 2012, \aap, 538, A19
\bibitem[Kalogera et al.(2004)]{kalogera2004} Kalogera, V., King, A.~R., \& Rasio, F.~A.\ 2004, \apjl, 601, L171
\bibitem[Katz(1975)]{katz1975} Katz, J. I. 1975, \nat, 253, 698
\bibitem[Kim et al.(2007)]{kim2007} Kim, M., Wilkes, B. J., Kim, D.-W., et al. 2007, \apj, 659, 29
\bibitem[Kremer et al.(2018a)]{kremer2018a} Kremer, K., Chatterjee, S., Rodriguez, C.~L., et al.\ 2018a, \apj, 852, 29
\bibitem[Kremer et al.(2018b)]{kremer2018b} Kremer, K., Ye, C.~S., Chatterjee, S., et al.\ 2018b, \apjl, 855, L15
\bibitem[Kremer et al.(2018c)]{kremer2018c} Kremer, K., Chatterjee, S., Breivik, K., et al.\ 2018, \prl, 120, 191103
\bibitem[Kremer et al.(2019)]{kremer2019} Kremer, K., Chatterjee, S., Ye, C.~S., et al.\ 2019, \apj, 871, 38
\bibitem[Kremer et al.(2020a)]{kremer2020a} Kremer, K., Ye, C.~S., Chatterjee, S., et al.\ 2020a, in IAU Symposium. 351, Star Clusters: From the Milky Way to the Early Universe, ed. A. Bragaglia et al. (Cambridge: Cambridge Univ. Press), 357
\bibitem[Kremer et al.(2020b)]{kremer2020b} Kremer, K., Ye, C.~S., Rui, N.~Z., et al.\ 2020b, \apjs, 247, 48
\bibitem[Kulkarni et al.(1993)]{kulkarni1993} Kulkarni, S.~R., Hut, P., \& McMillan, S.\ 1993, \nat, 364, 421
\bibitem[Lee et al.(1999)]{lee1999} Lee, Y.-W., Joo, J.-M., Sohn, Y.-J., et al.\ 1999, \nat, 402, 55
\bibitem[Maccarone et al.(2007)]{maccarone2007} Maccarone, T.~J., Kundu, A., Zepf, S.~E., et al.\ 2007, \nat, 445, 183
\bibitem[Mackey et al.(2007)]{mackey2007} Mackey, A.~D., Wilkinson, M.~I., Davies, M.~B., et al.\ 2007, \mnras, 379, L40 
\bibitem[Mackey et al.(2008)]{mackey2008} Mackey, A.~D., Wilkinson, M.~I., Davies, M.~B., et al.\ 2008, \mnras, 386, 65
\bibitem[Majewski et al.(2000)]{majewski2000} Majewski, S.~R., Patterson, R.~J., Dinescu, D.~I., et al.\ 2000, Liege International Astrophysical Colloquia, 35, 619
\bibitem[Merritt et al.(2004)]{merritt2004} Merritt, D., Piatek, S., Portegies Zwart, S., et al.\ 2004, \apjl, 608, L25
\bibitem[Miller \& Hamilton(2002)]{miller2002} Miller, M.~C., \& Hamilton, D.~P.\ 2002, \mnras, 330, 232
\bibitem[Miller-Jones et al.(2015)]{millerjonse2015} Miller-Jones, J.~C.~A., Strader, J., Heinke, C.~O., et al.\ 2015, \mnras, 453, 3918
\bibitem[Morscher et al.(2013)]{morscher2013} Morscher, M., Umbreit, S., Farr, W.~M., \& Rasio, F.~A.\ 2013, \apjl, 763, L15 
\bibitem[Morscher et al.(2015)]{morscher2015} Morscher, M., Pattabiraman, B., Rodriguez, C., Rasio, F.~A., \& Umbreit, S.\ 2015, \apj, 800, 9
\bibitem[Myeong et al.(2018)]{myeong2018} Myeong, G.~C., Evans, N.~W., Belokurov, V., et al.\ 2018, \mnras, 478, 5449 
\bibitem[Noyola et al.(2008)]{noyola2008} Noyola, E., Gebhardt, K., \& Bergmann, M.\ 2008, \apj, 676, 1008
\bibitem[Peuten et al.(2016)]{peuten2016} Peuten, M., Zocchi, A., Gieles, M., et al.\ 2016, \mnras, 462, 2333
\bibitem[Pooley et al.(2003)]{pooley2003} Pooley, D., et al. 2003, \apj, 591, 131
\bibitem[Portegies Zwart \& McMillan(2000)]{portegies2000} Portegies Zwart, S.~F., \& McMillan, S.~L.~W.\ 2000, \apjl, 528, L17 
\bibitem[Ransom (2008)]{ransom2008} Ransom, S. M. 2008, in AIP Conf. Ser. 983, 40 Years of Pulsars: Millisecond Pulsars, Magnetars and More (Melville, NY: AIP), 415
\bibitem[Rasio et al.(2000)]{rasio2000} Rasio, F.~A., Pfahl, E.~D., \& Rappaport, S.\ 2000, \apjl, 532, L47 
\bibitem[Rodriguez et al.(2015)]{rodriguez2015} Rodriguez, C.~L., Morscher, M., Pattabiraman, B., et al.\ 2015, \prl, 115, 051101
\bibitem[Rodriguez et al.(2016)]{rodriguez2016} Rodriguez, C.~L., Chatterjee, S., \& Rasio, F.~A.\ 2016, \prd, 93, 084029
\bibitem[Samsing et al.(2014)]{samsing2014} Samsing, J., MacLeod, M., \& Ramirez-Ruiz, E.\ 2014, \apj, 784, 71
\bibitem[Shara, \& Hurley(2006)]{shara2006} Shara, M.~M., \& Hurley, J.~R.\ 2006, \apj, 646, 464
\bibitem[Shishkovsky et al.(2018)]{shishkovsky2018} Shishkovsky, L., Strader, J., Chomiuk, L., et al.\ 2018, \apj, 855, 55
\bibitem[Sigurdsson \& Hernquist(1993)]{sigurdsson1993} Sigurdsson, S., \& Hernquist, L.\ 1993, \nat, 364, 423
\bibitem[Smith et al.(2000)]{smith2000} Smith, V.~V., Suntzeff, N.~B., Cunha, K., et al.\ 2000, \aj, 119, 1239
\bibitem[Sollima et al.(2007)]{sollima2007} Sollima, A., Ferraro, F.~R., \& Bellazzini, M.\ 2007, \mnras, 381, 1575
\bibitem[Spitzer(1969)]{spitzer1969} Spitzer, L.\ 1969, \apjl, 158, L139
\bibitem[Strader et al.(2012)]{strader2012} Strader, J., Chomiuk, L., Maccarone, T.~J., et al.\ 2012, \nat, 490, 71 
\bibitem[Trenti et al.(2007)]{trenti2007} Trenti, M., Ardi, E., Mineshige, S., et al.\ 2007, \mnras, 374, 857
\bibitem[Trenti \& van der Marel(2013)]{trenti2013} Trenti, M., \& van der Marel, R.\ 2013, \mnras, 435, 3272
\bibitem[van der Marel \& Anderson(2010)]{vandermarel2010} van der Marel, R.~P., \& Anderson, J.\ 2010, \apj, 710, 1063
\bibitem[Vasiliev(2019)]{vasiliev2019} Vasiliev, E.\ 2019, \mnras, 484, 2832
\bibitem[Verbunt \& Hut(1987)]{verbunt1987} Verbunt, F., \& Hut, P.\ 1987, IAU Symp. 125, The Origin and Evolution of Neutron Stars, ed. D. J. Helfand \& J. H. Huang (Dordrecht: Reidel), 187
\bibitem[Wang et al.(2016)]{wang2016} Wang, L., Spurzem, R., Aarseth, S., et al.\ 2016, \mnras, 458, 1450
\bibitem[Watkins et al.(2020)]{watkins2020} Watkins, L.~L., van der Marel, R.~P., Bellini, A., et al.\ 2020, in IAU Symposium. 351, Star Clusters: From the Milky Way to the Early Universe, ed. A. Bragaglia et al. (Cambridge: Cambridge Univ. Press), 544
\bibitem[Weatherford et al.(2018)]{weatherford2018} Weatherford, N.~C., Chatterjee, S., Rodriguez, C.~L., et al.\ 2018, \apj, 864, 13
\bibitem[Weatherford et al.(2020)]{weatherford2020} Weatherford, N.~C., Chatterjee, S., Kremer, K., et al.\ 2020, \apj, 898, 162
\bibitem[Ye et al.(2020)]{ye2020} Ye, C.~S., Fong, W.-. fai ., Kremer, K., et al.\ 2020, \apjl, 888, L10
\bibitem[Ye et al.(2019)]{ye2019} Ye, C.~S., Kremer, K., Chatterjee, S., et al.\ 2019, \apj, 877, 122
\bibitem[Zaris et al.(2020)]{zaris2020} Zaris, J., Veske, D., Samsing, J., et al.\ 2020, \apjl, 894, L9
\bibitem[Zhao et al.(2020)]{zhao2020} Zhao, Y., Heinke, C.~O., Tudor, V., et al.\ 2020, \mnras, 493, 6033
\bibitem[Zocchi et al.(2019)]{zocchi2019} Zocchi, A., Gieles, M., \& H{\'e}nault-Brunet, V.\ 2019, \mnras, 482, 4713

\end{thebibliography}
\end{document}